\documentclass[12pt]{article}
\usepackage{latexsym,cite,amssymb}
\usepackage{times}
\usepackage[dvipdf]{graphics}
\usepackage{color}
\usepackage{amsfonts}
\usepackage{latexsym,cite,amssymb}
\usepackage{times}
\usepackage{graphicx}
\usepackage{color}
\def\be{\begin{equation}}\def\ba{\begin{eqnarray}}
\def\ee{\end{equation}}\def\ea{\end{eqnarray}}

\makeatletter
\renewcommand{\theequation}{\thesection.\arabic{equation}}
\@addtoreset{equation}{section} \makeatother
\input amssym.def
\input amssym.tex

\begin{document}

\title{\vskip -60pt
\vskip 20pt Acoustic black holes from supercurrent tunneling~}
\author{ Xian-Hui Ge${}^{1,2}~$, Shao-Feng Wu${}^{1}~$, Yunping
Wang${}^{2}~$,\\
 Guo-Hong Yang${}^{1}~$ and  You-Gen Shen${}^{3}~$}
\date{}
\maketitle \vspace{-1.0cm}
\begin{center}
~~~
${}^{1}$Department of Physics, Shanghai University, Shanghai 200444, China\\
~{}
${}^{2}$Kavli Institute for Theoretical Physics China, CAS, Beijing 100190, China\\
${}^{3}$Institute of Physics, Chinese Academy of Sciences, Beijing 100190, China\\
~{}
${}^{4}$Shanghai Astronomical Observatory, Chinese Academy of Sciences, Shanghai 200030, China\\
~{}
\\

~~~\\
~~~\\
\end{center}

\begin{abstract}
We present a version of acoustic black holes by using the principle
of the Josephson effect. We find that in the case two superconductors $A$ and $B$ are
separated by an insulating barrier, an acoustic black hole may be
created in the middle region between the two superconductors. We
discuss in detail how to describe an acoustic black hole in the
Josephson junction and write the metric in  the langauge of the
superconducting electronics. Our final results infer that for big
enough tunneling current and  thickness of the junction,
experimental verification of the Hawking temperature could be
possible.
 \end{abstract}

{\small
\begin{flushleft}
\end{flushleft}}
\newpage

\tableofcontents 

\section{Introduction }
Analog models of gravity have recently received great interests
since these models may provide possible experimental verifications
of the evaporation of black holes. Unruh was the first to propose
the idea of using hydrodynamic  flows as analogues to mimic some
properties of black hole physics\cite{unruh} (for reviews see \cite{visser} and references there in). Any moving
fluid with speed exceeding the local sound velocity through a
spherical surface could in principle form an acoustic black hole.
For acoustic black holes, it is sound waves instead of light waves that
cannot escape from the horizon where the horizon locates  on
the boundary between subsonic and supersonic flow regions. In
particular, superfluid helium II \cite{volv}, atomic Bose-Einstein
condensates \cite{garay,garay1,barcelo,vis}, one-dimensional Fermi-degenerate
noninteracting gas \cite{gio} were proposed to create an
acoustic black hole geometry in the laboratory.

Because the Hawking temperature depends on the gradients of the flow
speed at the horizon,  detecting thermal phonons radiating from the
horizons is very difficult. In fact the Hawking temperature
calculated from models in Bose-Einstein condensates so far is very
low ($\sim$ nano Kelvin). Up to now, only few experimenters have
claimed that acoustic black holes are able to be observed. The very
recent experimental realization of acoustic black hole reported was
conducted in a Bose-Einstein condensate \cite{laha}. Considering that the
Bose-Einstein condensate systems have a very strict requirement on
the environment temperature (for example 170-nano Kelvin for the gas
of Rubidium atoms), the authors in \cite{gs} proposed that acoustic
black holes may be realized in superconducting materials which have
much higher critical temperatures ($\sim$ 100 Kelvin) and a
relativistic version of acoustic black holes was presented
there (see also\cite{bilic,fag,paris,ana,g,g1} for further reading).
It was observed that an acoustic black hole may form near the spiral  vortex
core in a type II superconductor \cite{gs}. But the experimental detecting of
such behavior in superconductors is so far very difficult, because
the superconducting phase could be destroyed when the speed of the
current carriers exceeds the Fermi-velocity. The purpose of this
paper is to investigate the possibility of creating acoustic black
holes in the supercurrent tunneling, in particular the Josephson
junctions, because there are some very sensitive detectors based on
the principle of the Josephson effect, for example, the
superconducting quantum interference device (SQUID).

\section{Brief reviews on the Josephson effect}
In the theory of superconductivity, the Ginzburg-Landau equation is a phenomenologically based theory, which has been proven to be very successful partly because it can describe the mixed states in type II superconductors. It was later shown by Gorkov that this equation can be derived from
the Bardeen-Cooper-Schrieffer (BCS) theory  in the case of thermodynamic equilibrium and for temperatures
close to the transition temperature \cite{gork}. The generalization of the Ginzburg-Landau equation, which includes the time dependent term, allows the superconductor to relax in its equilibrium state.
Now we consider two superconductors, A and B, separated by an insulating
barrier.  If the barrier is thick enough so that the superconductors
are isolated from each other, the time-dependent Ginzburg-Landau
equation for each side is \ba i\hbar
\partial_{t}\psi_A=H_A\psi_A+\kappa\psi_B, \label{h1}\\
i\hbar\partial_{t}\psi_B=H_2\psi_B+\kappa \psi_A, \label{h2}\ea
where \ba
H_A=\bigg[-\frac{\hbar^2}{2m_A}\left(\nabla+\frac{2ie}{\hbar}\vec{A}\right)^2+a_A(T)+b_{A}(T)|\psi_A|^2
\bigg],\\
H_B=\bigg[-\frac{\hbar^2}{2m_B}\left(\nabla+\frac{2ie}{\hbar}\vec{A}\right)^2+a_B(T)+b_{B}(T)|\psi_B|^2\bigg].
\ea $m_{A,B}$ is the mass of each cooper pair, $a_{A,B}(T)$ and
$b_{A,B}(T)$ are two parameters that depend on the temperature,
$\kappa$ is the coupling constant for the wave-functions across the
barrier. It is worth noting that $a_{A,B}(T)$ and
$b_{A,B}(T)$ are phenomenological parameters that can be fixed in experiments. Without the coupling $\kappa$,
no mass term is generated in our paper. We will set $m_A=m_B$ in the following.  Actually,
equations (\ref{h1}) and (\ref{h2}) are used to describe the
Josephson effect when a voltage $V$ is applied between the two
superconductors and then one can replace the Hamiltonian with
$H_A=eV$ and $H_B=-eV$. We use the coupled time-dependent
Ginzburg-Landau equations in order to obtain the effective acoustic
metric. Note that the similar computation was done for a
two-species Bose-Einstein condensate by using two-component
time-dependent Gross-Pitaevskii equations \cite{vis}. One may regard $b_A$ and $b_B$ as self-interactions $U_{AA}$ and $U_{BB}$ in \cite{vis}. But different from the two component Bose-Einstein condensate, here the two superconductors $A$ and $B$ are separated by a thin film and the interactions between the two superconductors are very weak. So the interactions $b_{AB}$ and $b_{BA}$ can be neglected. After
obtaining the metric, we will ask what we can learn from the
Josephson effect for our understanding on the acoustic black hole
physics.

The two macroscopic wave function can be written in the form \ba
\psi_A=\sqrt{\rho_A}e^{i\theta_A},\\
\psi_B=\sqrt{\rho_B}e^{i\theta_B}.
 \ea
 The equations of motion then becomes
 \ba
&&\partial_{t}\rho_i+\frac{\hbar}{m_{i}}\nabla\cdot(\rho_i
\vec{v}_i)+\frac{\kappa}{\hbar}\sqrt{{\rho_i}{\rho_j}}\sin(\theta_i-\theta_j)=0,\label{rea}\\
&&\hbar\partial_{t}\theta_i=\frac{\hbar^2}{2 m_i}\frac{\nabla^2
\sqrt{\rho_i}}{\sqrt{\rho_i}}-\frac{m_i}{2}\vec{v}^2_i-a_i(T)-b_i(T)\rho_i-
\kappa\sqrt{\frac{\rho_j}{\rho_i}}\cos(\theta_j-\theta_i),\label{th}
 \ea
 where $\vec{v}_i=\frac{\hbar
\nabla\theta_i}{m_{i}}-\frac{2e}{m_i}\vec{A}$ and $i,j=A, B$ ($i
\neq j$).  The first term in the right hand of (\ref{th})
corresponds to the quantum potential. The above two equations are
completely equivalent to the hydrodynamic  equations for
 irrotational and inviscid fluid apart from the quantum potential and
the $\kappa$ term. In the long-wavelength approximation, the
contribution coming from the linearization of the quantum potential
can be neglected.   Since the current  densities $\rho_i$ in the superconductor $A$ is not  much different from that in the superconductor $B$, the quantum potential can be neglected in our derivation of the acoustic metric. The physics of the Josephson current is mainly described by the $\kappa$ term in equations (\ref{rea}) and (\ref{th}) \cite{poole}.  The Josephson relations for the pair density can
be obtained from (\ref{rea}) \ba &&\partial_{t}(\rho_A-\rho_B)
+\frac{\hbar}{m_{A}}\nabla\cdot(\rho_A\vec{v}_A-\rho_B\vec{v}_B)
+\frac{2\kappa}{\hbar}\sqrt{{\rho_A}{\rho_B}}\sin(\theta_A-\theta_B)=0.\label{29}\ea
It can be reduced to the standard Josephson relation when the second
term is dropped out, that is to say, \ba \label{j}&&
j=2e\partial_{t}(\rho_A-\rho_B)=
\frac{4e\kappa}{\hbar}\sqrt{{\rho_A}{\rho_B}}\sin(\theta_B-\theta_A)=j_c\sin\theta,\ea
where \be \label{jc}
j_c=\frac{4e\kappa}{\hbar}\sqrt{{\rho_A}{\rho_B}},~~~
\theta=\theta_B-\theta_A .\ee

   Actually, only in the Josephson junctions,
the speed of electrons cannot be neglected. The kinetic energy of
one cooper pair tunneled from one side to the other side changes
little. That is why one can replace the Hamiltonian with $H_A=eV$
and $H_B=-eV$ and ignore the kinetic terms of the Hamiltonian in
(\ref{th}). Following this procedure, we obtain the second Josephson
equation \be \frac{\partial (\theta_{B}-\theta_{A})}{\partial
t}=\frac{2eV}{\hbar}. \ee
In the insulating barrier, the density of
the current carriers $\rho_j$ is much lower than  the density of current carriers
in the superconductors $\rho_s$. For those insulating films who share the
same cross-sectional area with the superconductors on each side,
they share the same current. That is to say
\be
\rho_j v_j e S_1=\rho_s v_s e S_2,
\ee
where $S_1=S_2=S$ is the cross-sectional area,  $v_j$ denotes the speed of the current
carriers in the film (junction), and $v_s$ the speed in the superconductors.
 Therefore, the speed of the current
carriers in the film (junction) must be much bigger than their speed
in the bulk:\be v_j \gg v_s .\ee
 On the other hand, if the two superconductors, $A$ and
$B$, are not separated by an insulating barrier, but by the same
superconducting material with a cross-sectional area much smaller
than the cross-sectional area of each side (i.e. $S_1\ll S_2$), this is another kind of
Josephson junctions. In this case, the  density of current is  same
everywhere in the around circle (i.e. $\rho_j = \rho_s$). Thus, we still have
$
v_j \gg v_s $.  These can  justify why we can drop the
second term in (\ref{29}) and kinetic energy terms in (\ref{th})
when we derive the Josephson relations and how it is possible for
the creation of an acoustic black hole in the Josephson junction because the speed of the current carriers can be so fast that it may exceed the local speed of  sound. In
this paper, we will construct an acoustic black hole from the
supercurrent tunneling by considering the linearized perturbations
of equations (\ref{rea}) and (\ref{th}).

\section{The acoustic black hole metric}
 Now let us consider a fixed background
$(\rho_{i0},\theta_{i0})$ with small perturbations
$\rho_i=\rho_{i0}+\rho_{i1}$ and $\theta=\theta_{i0}+\theta_{i1}$ (
$i=A,B$ ). The leading order equations for $(\rho_{i0},\theta_{i0})$
can be written as \ba
&&\partial_t\rho_{A0}+\frac{\hbar}{m_A}\nabla\cdot(\rho_{A0}
\vec{v}_{A0})+\frac{\kappa}{\hbar}\sqrt{{\rho_{A0}}{\rho_{B0}}}\sin(\theta_{A0}-\theta_{B0})=0,\\
&&\hbar\partial_{t}\theta_{A0}=-\frac{m_A}{2}\vec{v}^2_{A0}-a_{A}(T)-b_{A}(T)\rho_{A0}-
\kappa\sqrt{\frac{\rho_{B0}}{\rho_{A0}}}\cos(\theta_{B0}-\theta_{A0}).\\
&&\partial_t\rho_{B0}+\frac{\hbar}{m_B}\nabla\cdot(\rho_{B0}
\vec{v}_{B0})+\frac{\kappa}{\hbar}\sqrt{{\rho_{B0}}{\rho_{A0}}}\sin(\theta_{B0}-\theta_{A0})=0,\\
&&\hbar\partial_{t}\theta_{B0}=-\frac{m_B}{2}\vec{v}^2_{B0}-a_{B}(T)-b_{B}(T)\rho_{B0}-
\kappa\sqrt{\frac{\rho_{A0}}{\rho_{B0}}}\cos(\theta_{B0}-\theta_{A0}).
\ea

Linearizing the equations (\ref{h1}) and (\ref{h2}), we obtain the
two coupled equations for the perturbation of the phases \ba
\frac{\partial \theta_{A1}}{\partial
t}+\vec{v}_{A0}\cdot\nabla\theta_{A1}=
-\frac{b_{A}\rho_{A1}}{\hbar}+
\frac{\kappa}{2\hbar}\frac{\rho_{A1}\sqrt{\rho_{B0}}}{\rho^{3/2}_{A0}}
-\frac{\kappa}{2\hbar}\frac{\rho_{B1}}{\sqrt{\rho_{A0}\rho_{B0}}},\\
\frac{\partial \theta_{B1}}{\partial
t}+\vec{v}_{B0}\cdot\nabla\theta_{B1}=
-\frac{b_{B}\rho_{B1}}{\hbar}+
\frac{\kappa}{2\hbar}\frac{\rho_{B1}\sqrt{\rho_{A0}}}{\rho^{3/2}_{B0}}
-\frac{\kappa}{2\hbar}\frac{\rho_{A1}}{\sqrt{\rho_{A0}\rho_{B0}}}.\ea
The coupled equations for the density perturbations are given by \ba
&&\partial_{t}\rho_{A1}+\nabla\cdot(\frac{\hbar}{m_A}\rho_{A0}\nabla
\theta_{A1}+\rho_{A1}\vec{v}_{A0})=\frac{2\kappa}{\hbar}
\sqrt{\rho_{A0}\rho_{B0}}(\theta_{B1}-\theta_{A1}),\\
&&\partial_{t}\rho_{B1}+\nabla\cdot(\frac{\hbar}{m_B}\rho_{B0}\nabla
\theta_{B1}+\rho_{B1}\vec{v}_{B0})=\frac{2\kappa}{\hbar}
\sqrt{\rho_{A0}\rho_{B0}}(\theta_{A1}-\theta_{B1}).\ea The above
equations governing the perturbation of phases and density are  very hard to decouple. We need impose some constraints
on the background parameters for the purpose of deriving the
acoustic metric. The superconductors $A$ and $B$ can be the same so
that $\rho_{A0}=\rho_{B0}=\rho_0,~ b_A=b_B=b$, $m_A=m_B$  and the
background phases also can be set to be equal ($\theta_{A0}=\theta_{B0}$). This implies that without perturbations
there are no currents crossing the junction and  background
velocities $\vec{v}_{A0}=\vec{v}_{B0}=\vec{v}_0$. But when the phase
$\theta$ is fluctuated, the supercurrent tunneling happens in the
junction and the background velocity $v_0$ can be regarded as the
function of the space variables $x_i$. In the region where $v_0$
exceeds the ``sound velocity" $c_s$, an acoustic black hole forms.
In the following, we will see how this can happen. The coupled
equations for phase and density perturbation can be written as \ba
\partial_{t}(\theta_{A1}-\theta_{B1})+\vec{v}_0\cdot\nabla(\theta_{A1}
-\theta_{B1})=-\frac{b}{\hbar}(\rho_{A1}-\rho_{B1})
+\frac{\kappa}{\hbar}\frac{\rho_{A1}-\rho_{B1}}{\rho_0},\label{theta}\\
\partial_{t}(\rho_{A1}-\rho_{B1})+\nabla\cdot\left[\frac{\hbar}{m}\rho_0\nabla(\theta_{A1}
-\theta_{B1})+(\rho_{A1}-\rho_{B1})\vec{v}_0\right]
=-\frac{4\kappa\rho_0}{\hbar}(\theta_{A1}-\theta_{B1}).\label{ro}\ea
It is convenient to introduce the notations \be
\theta_1=\theta_{A1}-\theta_{B1},~~~\rho_1=\rho_{A1}-\rho_{B1}, \ee
and\be \chi=\frac{b\rho_0-\kappa}{\hbar \rho_0}. \ee After
combining ($\ref{theta}$) and ($\ref{ro}$) as a single equation, we
have the wave equation for $\theta_1$ \ba
-\partial_{t}\left[\frac{1}{\chi}\left(\partial_{t}\theta_1+\vec{v}_0\cdot\nabla\theta_1\right)\right]
+\nabla\cdot\left[\frac{\hbar\rho_0}{m}\nabla\theta_1
-(\partial_{t}\theta_1+\vec{v}_0\cdot\nabla\theta_1)
\frac{\vec{v}_0}{\chi}\right]=-\frac{4\kappa\rho_0}{\hbar}
\theta_1. \ea The above equation is comparable with a massive
Klein-Gordon equation in curved space-time \be
\frac{1}{\sqrt{-g}}\partial_{\mu}(\sqrt{-g}g^{\mu\nu}\partial_{\nu}\theta_1)-\tilde{m}^2\theta_1=0.
\ee We can therefore read off the inverse acoustic metric
\begin{equation}
\label{inv}
 g^{\mu\nu}\equiv
\frac{\sqrt{m^3}}{\sqrt{\hbar^3\rho^3_0\chi}} \left[
\matrix{-1&\vdots&-v_0^j \cr
               \cdots\cdots&\cdot&\cdots\cdots\cdots\cdots\cr
           -v_0^i &\vdots&(c^2_s\delta^{ij}-v_0^{i}v_0^{j})  \cr }
\right],
\end{equation}
and \be
\tilde{m}^2=-\frac{4\kappa\rho_0}{\hbar}\sqrt{\frac{m^3\chi}{\hbar^3\rho^3_0}
},\ee where the local speed of sound is defined as \be\label{cs}
c^2_s=\frac{\hbar\rho_0}{m}\chi=\frac{b\rho_0-\kappa}{m}. \ee Note
that in absence of the coupling constant $\kappa$, the local speed
of sound has the form \be c_s=\frac{\hbar}{\sqrt{2}m\xi(T)}, \ee
where $\xi(T)$ is the Ginzburg-Landau coherence length
$\xi(T)=\frac{\hbar}{\sqrt{2m|a(T)|}}$.

 By inverting (\ref{inv}), we determine the metric
\begin{equation}
\label{un}
 g_{\mu\nu}\equiv
\left(\frac{\hbar\rho_0}{m c_s}\right) \left[
\matrix{-(c^2_s-v^2_0)&\vdots&-v^j_0 \cr
               \cdots\cdots&\cdot&\cdots\cdots\cdots\cdots\cr
           -v^i_0 &\vdots&\delta^{ij}  \cr }
\right].
\end{equation}
In the presence of an external magnetic field, we will show in the
following that the structure of the acoustic black hole may have
``draining bathtub'' form. The general acoustic metric is given by
\be \label{ac}
ds^2=\frac{\hbar\rho_0}{mc_s}\bigg[-(c^2_s-v^2_0)dt^2-2\vec{v_0}\cdot
d\vec{r}dt+d\vec{r}\cdot d\vec{r}\bigg], \ee
where the corresponding horizon locates at $c_s=v_0$.

In the cylindrical coordinate $(r,\theta,z)$, suppose that the
superconducting current is along the $z$-direction and the magnetic
field $\textbf{A}$ is along the $\theta$-direction. In this case,
the background fluid flow will be bended by the magnetic field.  It
is convenient to set $v_z(r,
\theta,z)\neq 0$ and $v_{\theta}(r,
\theta,z)\neq 0$ and let  $v_r=0$. The metric
(\ref{ac}) becomes \ba\label{bath}
&&ds^2=\bigg(\frac{\hbar\rho_0}{mc_s}\bigg)\bigg[-(c^2_s-v^2_0)dt^2-2\vec{v_0}\cdot
d\vec{r}dt+d\vec{r}\cdot d\vec{r}\bigg]\nonumber\\
&&= \bigg(\frac{\hbar\rho_0}{mc_s}\bigg)\bigg[-c^2_s
dt^2+(v_{\theta}dt-rd\theta)^2+(v_{z}dt-dz)^2+dr^2\bigg],\ea where
$v^2_0=v^2_z+v^2_{\theta}$. If we make the coordinate
transformations \ba
&&dt=d\tau-\frac{v_z}{c^2_s-v^2_z} dz,\\
&&d\theta=d\vartheta-\frac{v_{\theta}v_z}{r(c^2_s-v^2_z)}dz,\ea then
the line-elements of the metric can be written as \be
ds^2=\bigg(\frac{\hbar\rho_0}{mc_s}\bigg)\bigg\{-[c^2_s-(v^2_z+v^2_{\theta})]
d\tau^2-2rv_{\theta}d\tau
d\vartheta+\frac{c^2_s}{c^2_s-v^2_z}dz^2+r^2d\vartheta^2+dr^2\bigg\}.\ee

 The
formation of an acoustic black hole  requires $\frac{v^{i}_0}{c_s}>
1$ in some regions. In \cite{gs}, the authors pointed out that it
would be very difficult to realize acoustic black holes by using
type I and type II superconductors. Especially, for type I
superconductors, when the speed of superconducting electrons is
equivalent to  the ``sound velocity", say $v_0=c_s$, the
superconducting phase is broken and return to the normal state. It
was expected to form an acoustic black hole in the region of the spiral  vortex
core in a type II superconductor.  The calculation shows that
$\xi<r<\sqrt{2}\xi$, electron velocity may exceed the sound velocity
\cite{gs}, where $r$ denotes the distance from the vortex core and
$\xi$ the coherence length. But experimental verification of
superconducting electron speed in the bulk of a superconductor is
not easy. From ($\ref{jc}$) (i.e. $j_c=2ev^i_0\rho_0\backsimeq
4e\kappa \rho_0/\hbar$) and ($\ref{cs}$), we know that
$\frac{v^{i}_0}{c_s}> 1$ means \be\label{ca1} \frac{v^{i}_0}{c_s}=
\frac{2\kappa}{\hbar}\sqrt{\frac{m}{b\rho_0-\kappa}} >1.\ee
 As an example, let us consider the material $\rm PbCu$ at $T_c=7.2 K$ with coherence length  $\xi=80 nm$ \cite{poole}. Then (\ref{ca1}) requires the coupling constant
 $\kappa>1.788\times 10^{-18} J$ , which is possible in experiments.
The Josephson junctions and related instruments, such as
superconducting quantum interference device (SQUID), may open a door
to build an acoustic black hole directly.

\section{A microscopic picture}
From (\ref{cs}), we know that the local speed of sound in the
junction could be smaller than that in the bulk of the
superconductor for a positive-valued coupling constant $\kappa$. In
the above derivation, we have used the Ginzburg-Landau theory, which
is very useful in describing qualitative and macroscopic behaviors.
In order to have a clear picture for the formation of acoustic black
holes, now we present a microscopic description by using the
BCS theory. In fact, Gorkov in 1959
proved that the Ginzburg-Landau theory can be derived from  full the
BCS theory in a suitable limit\cite{gork}. The microscopic model can
be constructed by using the tunneling Hamiltonian \ba
&&H=H_{R}+H_{L}+H_{T},\\
&&H_{T}=\sum_{kp\sigma}
(T_{kp}C^{\dag}_{k\sigma}C_{p\sigma}+h.c.).\ea The Hamiltonian is
identical with equations (\ref{h1}) and (\ref{h2}). $H_R$ is the
Hamiltonian for particles on the right side of the tunneling
junction. Similarly, $H_L$ has all the physics for particles on the
left side of the junction. The tunneling is caused by the term
$H_{T}$ and $T_{p\sigma}$ denotes the tunneling matrix that can
transfer particles through an insulating junction. The derivation of
the tunneling current led to two terms: the single-particle terms
and the Josephson term.  We only consider the Josephson term here.
 \be
i\hbar\partial_{t}\psi_i=H_i\psi_i+\kappa\psi_j \label{1}, \ee where
\ba
&&H_i=-\frac{\hbar^2}{2m_i}\left(\nabla+\frac{2ie}{\hbar}\vec{A}\right)^2
+\frac{1}{\eta}\left[\frac{T_c-T}{T_c}-\frac{1}{\rho_0}|\psi_i|^2\right],\nonumber\\
&&\eta=\frac{7\zeta(3)}{6(\pi T_c)^2}\varepsilon_{F}, \ea where
$\varepsilon_{F}$ is the Fermi energy. In this sense, the formation
of an acoustic black hole requires  \be\label{ca} \frac{v^{i}_0}{c_s}=
\frac{2\kappa}{\hbar}\sqrt{\frac{m\eta}{1-\eta\kappa}}
>1.\ee
As pointed out in section 2, in the insulating barrier, the density of the current carriers is much lower than that in the superconductors, but the speed of the current carriers in the film should be much bigger than their
speed in the bulk.  Therefore, there may exist a region in where $v_0<c_s$ that continuously connected to the region $v_0>c_s$.
\section{An acoustic black hole in the Josephson junction}
Let us consider a weak link tunnel junction with a magnetic field
$B_x (y) \vec{{i}}$ applied along the x-direction, as shown in
Fig.1. The junction is of thickness $2a$ normal to the $z$-axis with
cross-sectional dimensions $d$ and $w$ along $y$ and $x$,
respectively. We assume that the external magnetic field is larger
than the field produced by the currents. The applied field is
derived from the vector potential $\vec{\textbf{A}}=B_x(y)\vec{{k}}$. In the barrier film  the
material is normal and the magnetic field  is constant-valued $B_x(y)= B_0$, but the magnetic field decays exponentially into the
superconductors on either side of the junction.

 From (\ref{bath}), we know that the magnetic field can change the
direction of the fluid flow. The formation of an acoustic black hole
should satisfy the condition $v^{i}_0> c_s$. Note that when we
derive the acoustic metric,  we consider the perturbations around a
fixed background $(\rho_{i0},\theta_{i0})$ without the fluctuations
of the magnetic field. The vector potential $\vec{\textbf{A}}$ is regarded as an
external source.  We know that the phase $\theta^i_0$ is determined
by the magnetic field and the supercurrent. Let us first review the derivation of the
\textit{Josephson junction diffraction equation}( see \cite{wu,poole}
for more details).
\begin{figure}[htbp]
 \begin{minipage}{1\hsize}
\begin{center}
\includegraphics*[scale=0.6] {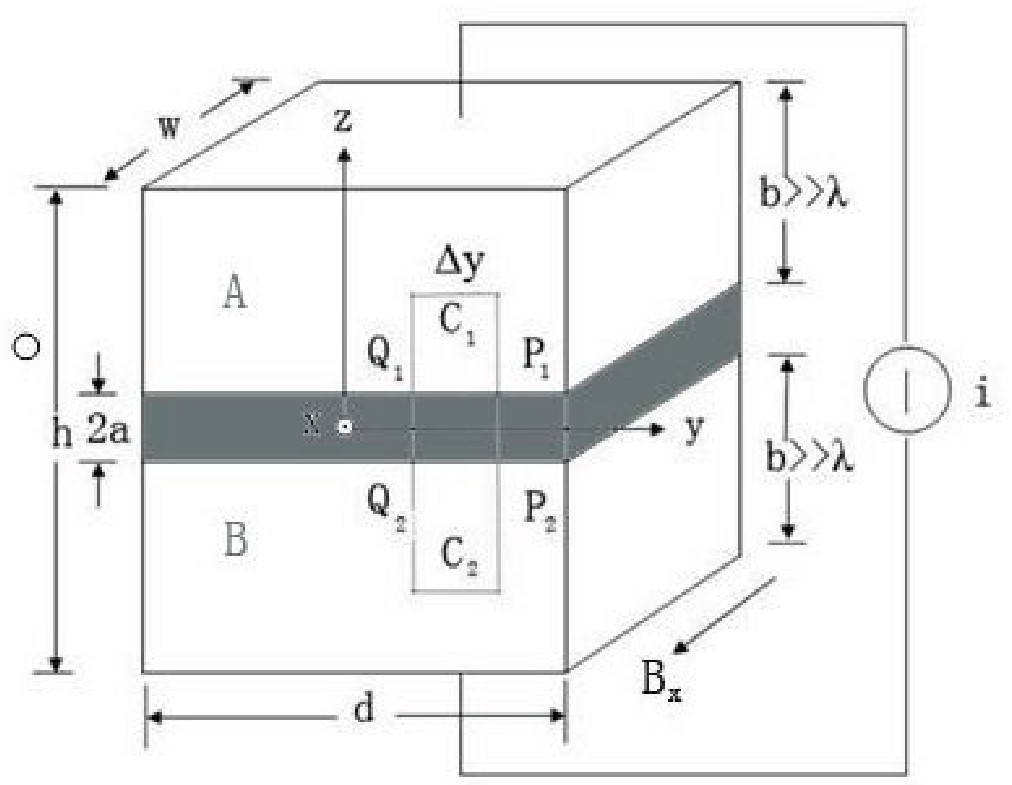}
\end{center}
\caption{ Application of a magnetic field $B_x$ transverse to the
Josephson junction. An acoustic black hole may be created in the
junction.} \label{B1}
\end{minipage}
\end{figure}
Consider a rectangle circle $PC_1QC_2P$ in the Josephson junction,
where $P$ and $Q$ locate at the middle of the junction. We can neglect the
thickness of the film and assume $\bigtriangleup y$ deep inside the superconductor where the
induced current decays away (i.e. the magnetic field is vanishing). The integration of $ \textbf{v}_z \cdot
d \textbf{l}$ along the  rectangle circle $PC_1QC_2P$  is zero.
Therefore, only the magnetic field contributes to the change of the
phase \be \nabla \theta^i_0=\frac{2e}{\hbar} \vec{A}.\ee In the
superconductor A, the integral around $C_1$ path gives \be
\theta_{0Q_1}(y)- \theta_{0P_1}(y+\Delta
y)=\frac{2e}{\hbar}\int_{C_1}\vec{A}\cdot d\vec{l}.\ee For the $C_2$
path \be \theta_{0P_2}(y)- \theta_{0Q_2}(y+\Delta
y)=\frac{2e}{\hbar}\int_{C_2}\vec{A}\cdot d\vec{l}. \ee Then we have
\be \theta_0(y+\Delta y)-
\theta_0(y)=\frac{2e}{\hbar}\oint\vec{A}\cdot d\vec{l}.\ee By using
the Stokes theorem, we find that \be\label{ma}
\frac{\partial\theta_0}{\partial y}\Delta y=\frac{2e}{\hbar}
\int\int\vec{B}\cdot
d\vec{s}=\frac{2e}{\hbar}B_{x}(2\kappa+2a)\Delta y.\ee Now, we have
\be \label{56}\nabla_y \theta_0= \frac{2e\kappa}{\hbar}B_x(y),\ee where
$\Lambda=2\kappa_L+2a$ denotes the effective thickness of the
junction and $\kappa_L$ is the penetration depth of the magnetic
field. The phase $\theta_0$ then depends on the coordinate $y$ \be
\theta_0(y)\simeq\frac{2e\Lambda B_0}{\hbar}y+c_1. \ee It is worth noting
that we can replace $B_x$ with the total flux across the junction
$\Phi_J=\Lambda dB_x$,  \be \frac{2e\Lambda
B_x}{\hbar}=\frac{2\pi\Lambda d
B_x}{\frac{\pi\hbar}{e}d}\approx\frac{2\pi\Phi_{J}}{d\phi_0}, \ee where
$\phi_0=\frac{h}{2e}$ is the magnetic quantum flux. If this is
substituted in Eq.(\ref{j}) and integrated over the area $S=w d$ of
the junction, we have \be I_s(B)=j_cS\frac{\sin(\frac{\pi
\Phi_J}{\phi_0})}{\frac{\pi \Phi_J}{\phi_0}} \sin c_1.\ee We call
this the \textit{Josephson junction diffraction equation}. This
equation indicates that the $n$th maximum of the current $I_s$
occurs at the flux value $\Phi_J=(n+\frac{1}{2})\phi_0$, but cancels
for $\Phi_J=n\phi_0$,  where $n$ is an integer.

We are interested in (\ref{56}), since this equation gives us
$v_y=\frac{\hbar}{m}\nabla_y \theta_0= \frac{2e\Lambda}{m}B_x(y)$.
Therefore, in the cartesian coordinate system, from (\ref{un}) we have the metric
\be\label{jm}
ds^2=\bigg(\frac{\hbar\rho_0}{mc_s}\bigg)\bigg[-(c^2_s-(v^2_z+v^2_y))dt^2-2{v_z}
dzdt-2{v_y} dydt+d\vec{x}_i\cdot d\vec{x}_i\bigg].\ee Taking the
coordinate transformation \ba
&&dt=d\tau-\frac{v_y}{c^2_s-v^2_y-v^2_z}
dy-\frac{v_z}{c^2_s-v^2_y-v^2_z}dz,\ea we obtain \ba \label{j2}
ds^2&=&\bigg(\frac{\hbar\rho_0}{mc_s}\bigg)\bigg[-[c^2_s-(v^2_z+v^2_y)]d\tau^2
+\left(\delta_{ij}+\frac{v^iv^j}{c^2_s-(v^2_z+v^2_y)}dx^idx^j\right)\bigg]
,\ea where we should note that
$c_s=\sqrt{\frac{b\rho_0-\kappa}{m}}$,
$v_z\approx\frac{2\kappa}{\hbar}$ and $v_y=\frac{2e\Lambda}{m}B_x(y)$. Note that the transformation from the metric (\ref{jm}) to (\ref{j2}) means that the resulting metric (\ref{jm})  is a static one. The variable
$t$ measures the time of the background fluid and $\tau$ is a redefined time.
It would be non-trivial to consider the case in which the total
phase change across the junction is $2n\pi$, with $n$ Josephson vortices
side by side in the junction, each containing one flux quantum. Then
the total current in the junction is vanishing because the
supercurrent flows down across the junction on the left and up on
the right. The current flows horizontally within a penetration depth
$\lambda$ inside the superconductor to form a closed loops. These
current loops encircle flux and the resulting configuration is known
as a Josephson vortex. The supercurrent  along the $z$-direction can
somehow be regarded as a constant value and then $v_z$ does not change
in the junction.

As a special condition, let us consider the Hawking temperature
contributed by $v_y$. In the vicinity of the horizon, we can split
up the fluid flow into normal and tangential components (i.e.
$\textbf{v}=\textbf{v}_z+\textbf{v}_y$) and we choose  $\hat{j}$ as
the unit vector field that at the horizon is perpendicular to
it\cite{visser}. From the definition, we know that the horizon
locates at $c_s={v}_y$.
  In this case,
the Hawking temperature at the event horizon is given by
\begin{equation}
T_{H}=\frac{\hbar}{2 \pi k_B }{\bigg|\partial_y
(c_s-v_{y})\bigg|}_{\rm horizon}.
\end{equation}
More explicitly, \ba {T}_{H}&=&\left(1.2\times 10^{-9} K
m\right)\left(\frac{1}{1000 m s^{-1}}\right)\bigg|\partial_y (c_s-v_{y})\bigg|_{\rm horizon}\nonumber\\
 &=&\left(1.2\times
10^{-9} K m\right)\left(\frac{1}{1000 m
s^{-1}}\right)\bigg|\left(\partial_y c_s-\frac{2e\Lambda \mu_0
j_z}{m}\right)\bigg|_{\rm horizon}, \ea where we have used the
Maxwell equation $\nabla \times \textbf{B}=\mu_0 \textbf{J}$. If the
speed of sound is to be a position-independent constant, the
resulting Hawking temperature then has the form \be
{T}_{H}=\left(1.2\times 10^{-9} K m\right)\left(\frac{1}{1000 m
s^{-1}}\right)\bigg|\frac{2e\Lambda \mu_0 j_z}{m}\bigg|_{\rm
horizon}.\ee This is a very interesting result which indicates that for
big enough value of the effective thickness of the junction
$\Lambda$ and the tunneling current $j_z$, the Hawking temperature
would be detectable in the future. As an example, let us estimate the Hawking temperature  of a particular kind of Josephson junction:  given the tunneling current $j_z=5\times 10^{7}$ $\rm A/m^2$
and the effective length $\Lambda=50 \rm nm$ (including the penetration depth)\cite{wu}, the resulting Hawking temperature is about $T_{H}\sim 10^{-7} \rm Kelvin$.
This value varies for different tunneling currents $j_z$. Compared with the Hawking temperature ($\sim$ nano Kelvin) of acoustic black holes in Bose-Einstein condensate, the temperature obtained here is two orders of magnitude higher that  maybe possible for the future experiments.

\section{Conclusion}
In summary, we have presented a version of acoustic black holes by
using the Josephson effect. We started from two coupled
Ginzburg-Landau equations with a coupling constant $\kappa$ for
Josephson junctions and reviewed the basic equations for the
Josephson effect.

The acoustic black hole metric can be obtained from the perturbation
equation for $\theta_1$ and $\rho_1$. The advantage of creating
acoustic metric by using the Josephson effect is that the coupling
constant $\kappa$ can be tuned. So that the sound velocity-- $c_s$
can be tuned to be very small and then $v_{0i}> c_s$ would become
easier. We discuss in detail how to describe an acoustic black hole
in the Josephson junction and write the metric in  the language of
the superconducting electronics. Finally, we estimate the Hawking
temperature of acoustic black hole created in the Josephson
junction. Although we have set up a theoretical model for acoustic black holes in Josephson junctions,  the experimental detection of the Hawking temperature would be difficult. Our result indicates that
the Hawking temperature strongly depends on the tunneling current. The enhancement of the Josephson current is thus crucial for the measurement of the acoustic black hole.

On the other hand, the obstacle of detecting Hawking radiation may come from its instability  against other mechanisms. For instance, for type I superconductors, when $v_0 = c_s$, the superconducting phase is broken and return to normal states \cite{gs}. In \cite{Bal}, the authors studied acoustic horizons in the quantum de Laval nozzle.  They  solved the Gross-Pitaevskii equation and found that  both in hydrodynamic and non-hydrodynamic regimes there
exist dynamically unstable regions associated with the creation of positive and negative energy quasiparticle pairs in analogy with the gravitational Hawking effect. In this paper, we may suffers the same problems since the Ginzburg-Landau equation and the Gross-Pitaevskii equation are very similar. Also, the quasinormal modes analysis of the obtained acoustic black holes may reveal that in the high momentum regime the configuration would be unstable against perturbations.  We leave discussion of the quantum instability to a future publication.

 \vspace*{10mm} \noindent
 {\large{\bf Acknowledgments}}

\vspace{1mm}  The work of XHG was partly supported by NSFC, China
(No. 10947116 and No. 11005072),  Shanghai Rising-Star Program and
SRF for ROCS, SEM. SFW was partly supported by NSFC, No.10905037.
YHG was partly supported by Shanghai Leading Academic
Discipline Project (No. S30105) and NSFC.
\renewcommand{\theequation}{A.\arabic{equation}}

\setcounter{equation}{0} \setcounter{footnote}{0}

\end{document}